\documentclass{aa}
\input{epsf}
\usepackage{graphicx}
\begin{document}
 
\title{SED, Metallicity and Age of Halo Globular Clusters in M33}

\author{Jun Ma, Xu Zhou, Jiansheng Chen, Hong Wu,
Zhaoji Jiang, Suijian Xue, Jin Zhu}

\offprints{Jun Ma, \\
\email{majun@vega.bac.pku.edu.cn}}
\institute{National Astronomical Observatories,
Chinese Academy of Sciences, Beijing, 100012, P. R. China}
\date{Received........; accepted........}


\newcommand\beq{\begin{equation}}
\newcommand\eeq{\end{equation}}
\newcommand\mum{\mu{m}}
\def\DeltaRA{|\Delta{\rm RA|}}
\def\DeltaDec{|\Delta{\rm DEC|}}


\abstract{
In this paper we study the properties of ten halo globular
clusters in the nearby spiral galaxy M33. CCD images of M33
were obtained as a part of the BATC Colour Survey of the sky
in 13 intermediate-band filters from 3800 to 10000{\AA}.
By aperture photometry, we obtain the spectral energy
distributions (SEDs) of these globular clusters.
We estimate the ages of our sample clusters by comparing
the photometry of each object with theoretical stellar
population synthesis models for different values of metallicity.
Our results suggest that eight of the ten sample halo globular clusters
have ``intermediate'' ages, i.e. between 1 and 8 Gyrs.
}
\titlerunning{SED, Metallicity and Age of Halo Globular Clusters in M33}
\authorrunning{Ma et al.}
\maketitle

\section{Introduction}

Globular clusters are roughly spherical agglomerations of stars.
It is generally believed that they are among
the first objects to be formed in a galaxy. Their ages provide
us with information on the early formative stages of the parent galaxy.
The study of these systems has played a key role in the development
of our understanding of the parent galaxy. For example, they can be
utilized to provide a lower limit to the age of the parent galaxy
provided their ages can be ascertained, and to study
the properties of the parent galaxy soon after its formation.

In addition to the Milky Way, a number of other Local Group galaxies
have been presented to
contain globular clusters: for example, the Sagittarius dwarf
spheroidal galaxy, the Large and Small Magellanic Clouds,
M31, and M33.
Schommer et al. (1991) estimated the number of the total ``true''
globular cluster population of M33 to be only $\sim 20$. Sarajedini
et al. (1998) selected ten halo globular clusters from Schommer et al. (1991)
by inspecting the difference between the cluster velocity and the
disk velocity as a function of the integrated cluster colour, and
constructed colour-magnitude diagrams to estimate the cluster
metallicity using the shape and colour of the red giant branch.
Under the assumption that cluster age is the global second
parameter, Sarajedini et al. (1998) presented that the average
age of halo globular clusters in M33 appears to be a few Gyrs younger
than halo clusters in the Milky Way.

In this paper, we present the SEDs of ten halo globular clusters of M33 from
Sarajedini et al. (1998),
and estimate the metallicities and ages of these clusters by using
the theoretical evolutionary
population synthesis methods (Bruzual \& Charlot 1996).

The outline of the paper is as follows.  Details of observations
and data reduction are given in Sect. 2. In Sect. 3, we provide
a brief description of the
stellar population synthesis models of Bruzual \&
Charlot (1996). The metallicities and ages
are estimated in Sect. 4.
The summary is presented in Sect. 5.

\section{Sample of star clusters, observations and data reduction}

\subsection{Sample of star clusters}

The sample of halo globular clusters in this paper is from
Sarajedini et al. (1998), who estimated their metallicities
using the shape and colour of the red giant branch.
The RAs and DECs of these clusters are from
Christian \& Schommer (1982), who detected more than
250 nonstellar objects (including the sample
halo globular clusters of this paper)
using $14\times 14$ $\rm inch^2$
unfiltered, unbaked, IIa-O focus plate exposed for
150 minutes with the Kitt Peak 4 m Richey-Chr\'{e}tien (R-C)
direct camera. Because of distortion in the 4 m
R-C focal plane, Christian \& Schommer (1982) presented the coordinates of
the sample clusters to be good only to $\sim 20''$.
We corrected the coordinates of our sample clusters  using
the HST Guide Star Catalog, and list them in Table 3.

\subsection{CCD image observation}

The large field multi-colour observations of the spiral galaxy M33 were
obtained in the BATC photometric system. The telescope used is the
60/90 cm f/3 Schmidt Telescope of Beijing Astronomical Observatory (BAO),
located at the Xinglong station. A Ford Aerospace 2048$\times$2048 CCD
camera with 15 $\mu$m pixel size is mounted at the Schmidt focus of the
telescope. The field of view of the CCD is $58^{\prime}$ $\times $ $
58^{\prime}$ with a pixel scale of $1\arcsec{\mbox{}\hspace{-0.15cm}.} 7$.  
The multi-colour BATC filter system includes 15 intermediate-band filters,
covering the total optical wavelength range from 3000 to 10000{\AA}
(see Fan et al. 1996). The images of M33 covering
the whole optical body of M33 were accumulated in 13 intermediate band
filters with a total exposure time of about 32.75 hours from September
23, 1995 to August 28, 2000.  The CCD images are centered at ${\rm
RA=01^h33^m50^s{\mbox{}\hspace{-0.13cm}.}58}$ and
DEC=30$^\circ39^{\prime}08^{\prime\prime}{\mbox{}\hspace{-0.15cm}.4}$
(J2000). The dome flat-field images were taken by using a diffuse plate in
front of the correcting plate of the Schmidt telescope. For flux calibration,
the Oke-Gunn primary flux standard stars HD19445, HD84937, BD+262606
and BD+174708 were observed during photometric nights. The parameters of
the filters and the statistics of the observations are given in Table 1.
Fig. 1 is the image of M33 in filter BATC07 (5785{\AA}), the circles
in which indicate the positions of the sample clusters in this paper.

\begin{figure*}
\hspace{-1.5cm}
\vspace{20cm}
\caption{The image of M33 in filter BATC07 (5785{\AA}) and the positions
of the sample globular clusters.
The center of the image is located at
${\rm RA=01^h33^m50^s{\mbox{}\hspace{-0.13cm}.}58}$
DEC=30$^\circ39^{\prime}08^{\prime\prime}{\mbox{}\hspace{-0.15cm}.4}$
(J2000.0). North is up and east is to the left.}
\end{figure*}

\subsection{Image data reduction }

The data were reduced with standard procedures, including bias subtraction
and flat-fielding of the CCD images, with an automatic data reduction
software named PIPELINE I developed for the BATC multi-colour sky survey
(\cite{Fan96}; \cite{Zheng99}).
The flat-fielded images of each colour were combined
by integer pixel shifting. The cosmic rays and bad pixels were corrected
by comparison of multiple images during combination. The images were
re-centered and position calibrated using the HST Guide Star Catalog.
The absolute flux of intermediate-band filter images was
calibrated using observations of standard stars. Fluxes as observed
through the BATC filters for the Oke-Gunn stars were derived by convolving
the SEDs of these stars with the measured BATC filter transmission
functions (\cite{Fan96}). {\it Column} 6 in Table 1 gives the zero point
error, in magnitude, for the standard stars in each filter. The formal
errors we obtain for these stars in the 13 BATC filters are $\la 0.02$
mag. This indicates that we can define the standard BATC system to an
accuracy of  $\la 0.02$ mag.

\subsection{Integrated photometry}

For each globular cluster, the PHOT routine in DAOPHOT (Stetson 1987)
was used to obtain
magnitudes. In order to avoid contamination from
nearby objects, a smaller aperture
of $6\arcsec{\mbox{}\hspace{-0.15cm}.} 8$, which corresponds
to a diameter of 4 pixels in Ford CCDs, was adopted.
Considering the large seeing of the Xinglong station (The typical
seeing is $2''$ in the Xinglong station.), aperture
corrections were computed using isolated stars.
Finally, the SEDs for ten globular clusters were obtained.
Table 2 contains the following information: {\it Column 1} is cluster
name which is taken from Christian \& Schommer (1982).
{\it Column} 2 to {\it Column} 14 show the magnitudes of
different bands. Second line of each star cluster is
the uncertainties of magnitude of corresponding band.
The uncertainties given for the magnitudes of each filter are
obtained from DAOPHOT's PHOT routine. For each cluster, the
background sky level was determined in an annulus of 
$\sim 14''-22''$.

\subsection{Comparison with previous photometry}

Using the Landolt standards, Zhou et al. (2001) presented the relationships
between the BATC intermediate-band system and $UBVRI$ broadband system
by the catalogs of Landolt (1983, 1992) and Galad\'\i-Enr\'\i quez et al. (2000).
We show the coefficients of the fits
in Eqs. (1) and (2).
\beq
m_B=m_{04}+(0.2218\pm0.033)(m_{03}-m_{05})+0.0741\pm0.033,
\eeq
\beq
m_V=m_{07}+(0.3233\pm0.019)(m_{06}-m_{08})+0.0590\pm0.010.
\eeq
By Eqs. (1) and (2), we transform the magnitudes of
ten halo globular clusters in BATC03, BATC04 and BATC05
bands to ones in $B$ band,
and in BATC06, BATC07 and BATC08 bands to ones in $V$ band.
Christian \& Schommer (1982) (hereafter CS1982) also obtained
photometry for these clusters. Fig. 2 plots the comparison of
$V$ (BATC) and ($B-V$) (BATC) photometry with previously
published measurements (Christian \& Schommer 1982), and
Table 3 shows this comparison.
The mean $V$ magnitude and colour differences (this paper's values minus
the values of Christian \& Schommer 1982) are
$<\Delta V> =-0.006\pm0.017$ and
$<\Delta (B-V)>=-0.125\pm0.040$, respectively.
The mean $V$ magnitude difference is very small,
but the mean colour difference is somewhat large.
Christian \& Schommer (1982) presented $BVR$ photometry for
60 star clusters in M33 using the KPNO
video camera mounted on the 2.1 m telescope.
They used NGC2264, NGC2419, NGC7006, and M15 for flux
calibration. Christian \& Schommer (1988)
presented $BVI$ photometry of 71 M33 star clusters
based on CCD frames obtained for 13 M33 fields.
They used the standard stars selected from the list
of Landolt (1983) for flux calibration.
By comparing the values of $B-V$ for the overlapping
star clusters in Christian \& Schommer (1982, 1988),
a large difference in the photometric measurements was found.
For H38, U50, and U102, the differences of $B-V$ are
larger than 0.15 (0.21, 0.17, and 0.24, respectively).
The remaining overlapping nine clusters have a mean colour difference
(the values of Christian \& Schommer
1988 minus the values of Christian \& Schommer 1982)
$<\Delta (B-V)>=-0.059\pm0.024$.
Table 4 shows the comparison of $BV$ colour photometry
for the overlapping star clusters in Christian \& Schommer
(1982, 1988), the colour differences of which are smaller than 0.15.
By analyzing the comparison above, we suggest that the
offset in colour between Christian \& Schommer
(1982) and this paper may be partly from the uncertainty of photometry
of Christian \& Schommer (1982), since Christian \& Schommer
(1982) only used NGC2264, NGC2419, NGC7006, and M15 for flux
calibration. The remaining offset may come from the uncertainty
of photometry in BATC03, BATC04 and BATC05 filters,
since there is about an uncertainty of 0.02 mag in the photometric measurements
of each filter of these three ones (see in Table 1).

\begin{figure*}
\hspace{0.5cm}
\vspace{-2.0cm}
\resizebox{15.cm}{!}{\rotatebox{-90}{\includegraphics{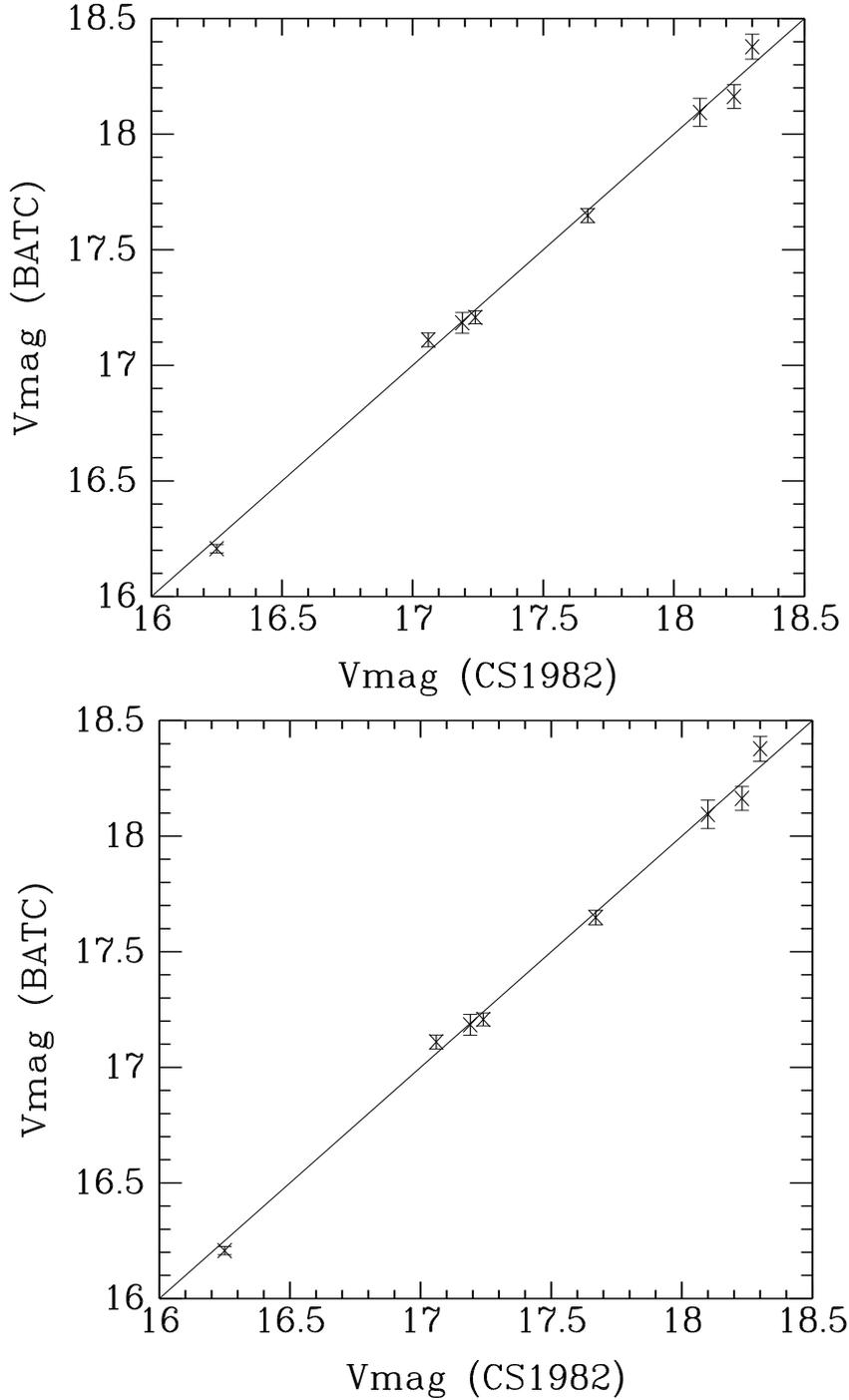}}}\\
\vspace{-2.0cm}
\hspace{0.5cm}
\resizebox{15.cm}{!}{\rotatebox{-90}{\includegraphics{H2904F2.1.ps}}}
\vspace{1cm}
\caption{Comparison of cluster photometry with previous measurements}
\end{figure*}

\section{Databases of simple stellar populations}

A simple stellar population (SSP) is defined to be a single generation
of coeval stars, which have fixed parameters such as metallicity, initial
mass function, etc. (\cite{Buzzoni97}).
SSPs are the basic building blocks of synthetic spectra
of galaxies that can be used to infer the formation and subsequent
evolution of the parent galaxies.
They are modeled by a collection of stellar evolutionary tracks with
different masses and initial chemical compositions, supplemented
with a library of stellar spectra for stars at different evolutionary
stages in evolution synthesis models (\cite{Jab96}).
In order to study the
integrated properties of halo globular clusters in M33,
we use the SSPs of Galaxy Isochrone Synthesis Spectra Evolution Library
(Bruzual \& Charlot 1996 hereafter GSSP). The main reason is that
the SSPs are simple and reasonably well understood.
 
\subsection{Spectral energy distribution of GSSPs}

Bruzual \& Charlot (1993) presented ``isochrone synthesis'' as a natural and
reliable approach to model the evolution of stellar
populations in star clusters and galaxies. With
this isochrone synthesis algorithm, Bruzual \& Charlot (1993)
computed the spectral energy distributions of stellar
populations with solar metallicity.
Bruzual \& Charlot (1996) improved the Bruzual \& Charlot
(1993) evolutionary population synthesis models. The updated version
provides the evolution of the spectrophotometric properties for a wide
range of stellar metallicity. They are based on the stellar evolution
tracks computed by Bressan et al. (1993), Fagotto et al.  (1994), and
by Girardi et al. (1996), who use the radiative opacities of Iglesias
et al. (1992). This library includes tracks for stars with metallicities
$Z=0.0004, 0.004, 0.008, 0.02, 0.05,$ and $0.1$, with the helium abundance
given by $Y=2.5Z+0.23$ (The reference solar metallicity is $Z_\odot=0.02$.).
The details can be seen from Ma et al. (2001).

\subsection{Integrated colours of GSSPs}

Kong et al. (2000) have obtained the age, metallicity, and interstellar-medium
reddening distribution for M81.
They found the best match between the intrinsic colours
and the predictions of GSSP for each cell of M81.
To determine the distributions of metallicity and age of the
star clusters in M33, we follow the method of Kong et al. (2000).
Since the observational
data are integrated luminosity, we need to convolve
the SED of GSSP with BATC filter profiles to obtain the optical
and near-infrared integrated luminosity for comparisons (Kong et al. 2000).
The integrated luminosity
$L_{\lambda_i}(t,Z)$ of the $i$th BATC filter can be calculated with
\beq
L_{\lambda_i}(t,Z) =\frac{\int
F_{\lambda}(t,Z)\varphi_i(\lambda)d\lambda} {\int
\varphi_i(\lambda)d\lambda},
\eeq
where $F_{\lambda}(t,Z)$ is the spectral energy distribution of
the GSSP of metallicity $Z$ at age $t$, $\varphi_i(\lambda)$ is the
response functions of the $i$th filter of the BATC filter system
($i=3, 4, \cdot\cdot\cdot, 15$),
respectively.
 
The absolute luminosity can be obtained if we know the distance to M33
and the extinction along the line of sight.
However, we do not know these parameters exactly. So, we should
work with the colours because of their independence of the distance.
We calculate the integrated colours of a GSSP relative to
the BATC filter BATC08 ($\lambda=6075${\AA}):
\beq
\label{colour}
C_{\lambda_i}(t,Z)={L_{\lambda_i}(t,Z)}/{L_{6075}(t,Z)}.
\eeq
As a result, we obtain intermediate-band colours for 6 metallicities from
$Z=0.0004$ to $Z=0.1$. 

\section{Reddening correction, metallicity and age of halo globular cluster}

In general, the SED of a stellar system depends on age, metallicity
and reddening along the line of sight. The effects of age, metallicity
and reddening are difficult to separate (e.g., \cite{Calzetti97};
\cite{Origlia99}; \cite{Vazdekis97}). Older age, higher metallicity or
larger reddening all lead to a redder SEDs of stellar systems in the
optical (\cite{Molla97}; \cite{Bressan96}). In order to obtain
intrinsic colours for these star clusters, we must correct for
reddening.

\subsection{Reddening correction}

The observed colours are affected by two sources of reddening: (1)
the foreground extinction in our Milky Way and (2) internal
reddening due to varying optical paths through the disk of the
parent galaxy.
Sarajedini (1994) presented the relation between the metal
abundance and reddening of a globular cluster, by which
the two parameters can be simultaneously derived using
the shape of the red giant branch (RGB), the $V$ magnitude of
the horizontal branch (HB), and the apparent $V-I$ colour of the
RGB at the level of the HB.
Using this relation, Sarajedini et al. (1998) obtained
the reddening (including both the foreground extinction in our Milky Way
and internal reddening) for nine clusters except for R14.
For R14, we may determine its reddening
by fitting the intrinsic colours and integrated colours of GSSP (see Sect 4.2).
The reddening is an important parameter in our paper, but
it is difficult to determine. As a first step, we only
adopted the values presented by Sarajedini et al. (1998).
They are included in Table 3.
We converted the values of ${\rm E}(V-I)$ to ${\rm E}(B-V)$ using a combination of
the relation $({\rm E}(V-I)/{\rm E}(B-V)=1.35)$ published by
Zinn \& Barnes (1996). Besides,
we adopted the extinction curve presented by Zombeck (1990).

\subsection{Metallicity and age distribution}

Since we model the stellar populations by SSPs, the intrinsic colours for
each star cluster are determined by two parameters: age, and metallicity.
In this section, we will determine these two parameters for these
star clusters simultaneously by a least square method.
The best fit age and metallicity
are found by minimizing the difference between the intrinsic colours and
integrated colours of GSSP:
\beq 
R^2(n,t,Z)=\sum_{i=3}^{15}[C_{\lambda_i}^{\rm
intr}(n)-C_{\lambda_i}^ {\rm ssp}(t, Z)]^2, 
\eeq
where $C_{\lambda_i}^{\rm ssp}(t, Z)$ represents the integrated colour in
the $i$th filter of a SSP with age $t$ and metallicity $Z$,
and $C_{\lambda_i}^{\rm intr}(n)$ is the intrinsic
integrated colour for nth star cluster.

Multi-colour photometry of each object was compared with
models for six different values of Z to determine ages for
globular clusters. In Table 5 (The results of R14
are listed in Table 6.), ages from each model are
given separately (We only list the models of four different values of Z.).
We also present the parameter $R^2$ of Eq. (5)
in Tables 5 and 6. The age of each cluster from the smallest $R^2$ is adopted, since this
is the best overall fit to each globular cluster.
Fig. 3 shows a mosaic of the fit of the integrated colour
of a SSP ($Z=0.0004, 0.004, 0.008$, and $0.02$)
with the intrinsic integrated colour for these ten globular clusters.
In Fig. 3, the thick line represents the best fit of the integrated
colour of a SSP of GSSP, and filled circle represents the intrinsic
integrated colour of a star cluster.
From Tables 5 and 6, we can see that
eight halo globular clusters
perhaps have ``intermediate'' ages, i.e.
between 1 and 8 Gyrs.
The results also that nine of the ten halo globular clusters
are metal poor.
Clusters C38 and U137 are as metal poor as $\rm{[Fe/H]}=-1.70$,
U77 is as metal rich as $\rm{[Fe/H]}=0.0$, respectively.
The other clusters are as metal poor as $\rm{[Fe/H]}=-0.70$.
In this paper, we may estimate ages of clusters that
have ``intermediate'' ages, however, cannot determine
metallicities of clusters since the models of GSSP
(Bruzual \& Charlot 1996) are not suited for a metallicity
determination. Although we presented the metallicity of each
cluster in Tables 5 and 6,
we only mean that, in this model of metallicity, the intrinsic integrated colour
of a cluster can do the best fit with the integrated colour of a SSP at some age. 
The results in this study may reflect possible non-astrophysical solutions.
Deep colour-magnitude diagrams would be necessary for conclusive  ages.
 
Since the value of reddening for globular cluster R14 is not presented by other
authors, we attempt to determine it and age simultaneously
by fitting the intrinsic
colours and integrated colours of GSSP using Eq. (5).
We select three models of metallicity (Z=0.02, 0.004, and 0.0004).
The results are listed in Table 6.

\begin{figure*}
\hspace{-5cm}
\resizebox{28.cm}{!}{\rotatebox{-90}{\includegraphics{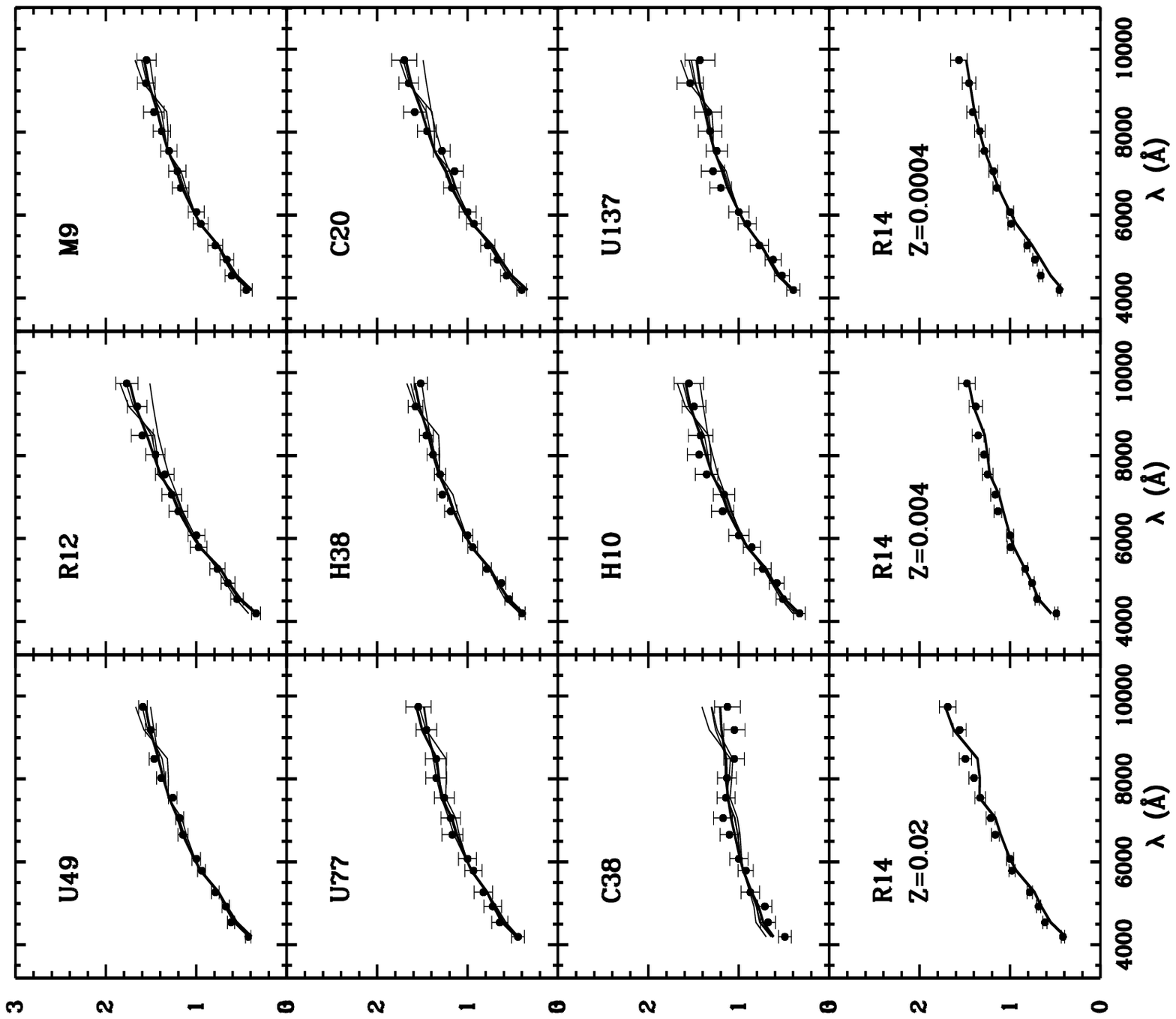}}}
\vspace{-1cm}
\caption{Mosaic of the fit of the integrated colour
of a SSP ($Z=0.0004, 0.004, 0.008$, and $0.02$)
with intrinsic integrated colour for sample halo globular clusters except for
R14, the intrinsic integrated colour of which is fitted by
three models of metallicity ($Z=0.02, 0.004$, and $0.0004$).
Thick line represents the best fit of the integrated colour of a SSP, and
filled circle represents the intrinsic integrated colour of a star cluster.}
\label{FigVibStab}
\end{figure*}

\section{Summary}

In this paper, we have, for the first time, obtained the SEDs
of ten halo globular clusters of M33 in 13 intermediate colours with
the BAO 60/90 cm Schmidt telescope.
Below, we summarize our main conclusions.
 
1. Using the images obtained with the Beijing Astronomical
Observatory $60/90$ cm Schmidt Telescope in 13 intermediate-band filters
from 3800 to 10000{\AA}, we obtained SEDs of ten halo globular clusters,
that were detected by Christian \& Schommer (1982),
and corrected their coordinates using the HST Guide Star Catalog.
 
2. Using theoretical stellar population synthesis
models, we obtained the distributions of
age for ten clusters. The results only suggest
that eight of the ten halo globular clusters
have ``intermediate'' ages, i.e.
between 1 and 8 Gyrs, however, these results
may reflect possible non-astrophysical solutions.

\begin{acknowledgements}
We would like to thank two anonymous referee for providing valuable comments
in the scientific contents and language of this paper.
We are greatly indebted to Dr. Habing for a very careful
reading of the manuscript and comments which improved the
presentation of our results.
We are grateful to the Padova group for providing us with a set of
theoretical isochrones and SSPs. We also thank G. Bruzual and
S. Charlot for sending us their latest calculations of SSPs and
for explanations of their code. The BATC Survey is supported by the
Chinese Academy of Sciences, the Chinese National Natural Science
Foundation and the Chinese State Committee of Sciences and
Technology. The project is also supported in part
by the National Science Foundation (grant INT 93-01805) and
by Arizona State University, the University of Arizona and Western
Connecticut State University.
\end{acknowledgements}


\begin{table*}
\caption[]{Parameters of the BATC filters and statistics of observations}
\begin{tabular}{cccccc}
\hline
\hline
 No. & Name& cw$^{\rm a}$(\AA)& Exp. (hr)&  N.img$^{\rm b}$
 & rms$^{\rm c}$ \\
\hline
1  & BATC03& 4210   & 00:55& 04 &0.024\\
2  & BATC04& 4546   & 01:05& 04 &0.023\\
3  & BATC05& 4872   & 03:55& 19 &0.017\\
4  & BATC06& 5250   & 03:19& 15 &0.006\\
5  & BATC07& 5785   & 04:38& 17 &0.011\\
6  & BATC08& 6075   & 01:26& 08 &0.016\\
7  & BATC09& 6710   & 01:09& 08 &0.006\\
8  & BATC10& 7010   & 01:41& 08 &0.005\\
9  & BATC11& 7530   & 02:07& 10 &0.017\\
10 & BATC12& 8000   & 03:00& 11 &0.003\\
11 & BATC13& 8510   & 03:15& 11 &0.005\\
12 & BATC14& 9170   & 01:15& 05 &0.011\\
13 & BATC15& 9720   & 05:00& 26 &0.009\\
\hline
\vspace{0.1cm}
\end{tabular}\\
$^{\rm a}$ Central wavelength for each BATC filter\\
$^{\rm b}$ Image numbers for each BATC filter\\
$^{\rm c}$ Zero point error, in magnitude, for each filter
as obtained\\
from the standard stars
\end{table*}

\begin{table*}
\caption{SEDs of ten halo globular clusters}
\begin{tabular}{cccccccccccccc}
\hline
\hline
 Cluster & 03  &  04 &  05 &  06 &  07 &  08 &  09 &  10 &  11 &  12 &  13 &  14 &  15\\
(1)    & (2) & (3) & (4) & (5) & (6) & (7) & (8) & (9) & (10) & (11) & (12) & (13) & (14)\\
\hline
     U49 &  16.99 &  16.56 &  16.46 &  16.27 &  16.06 &  15.98 &  15.82 &  15.78 &  15.70 &  15.58 &  15.52 &  15.48 &  15.40\\
 & 0.022 &  0.018 &  0.014 &  0.015 &  0.010 &  0.011 &  0.011 &  0.010 &  0.010 &  0.009 &  0.010 &  0.012 &  0.012    \\
     R12 &  17.34 &  16.81 &  16.62 &  16.43 &  16.15 &  16.11 &  15.90 &  15.83 &  15.77 &  15.68 &  15.56 &  15.51 &  15.44\\
&  0.049 &  0.036 &  0.027 &  0.029 &  0.020 &  0.021 &  0.022 &  0.023 &  0.022 &  0.019 &  0.021 &  0.020 &  0.024     \\
     R14 &  17.55 &  17.04 &  16.84 &  16.59 &  16.21 &  16.14 &  15.91 &  15.81 &  15.66 &  15.55 &  15.43 &  15.32 &  15.20\\
&  0.052 &  0.037 &  0.027 &  0.028 &  0.019 &  0.021 &  0.017 &  0.018 &  0.020 &  0.016 &  0.017 &  0.017 &  0.021     \\
     M9  &  17.82 &  17.48 &  17.38 &  17.17 &  16.96 &  16.90 &  16.73 &  16.68 &  16.60 &  16.53 &  16.45 &  16.39 &  16.38\\
&  0.046 &  0.034 &  0.025 &  0.026 &  0.016 &  0.018 &  0.016 &  0.018 &  0.018 &  0.017 &  0.025 &  0.021 &  0.026     \\
     U77 &  17.94 &  17.51 &  17.38 &  17.20 &  17.05 &  16.97 &  16.77 &  16.75 &  16.68 &  16.59 &  16.58 &  16.49 &  16.41\\
 & 0.066 &  0.047 &  0.042 &  0.039 &  0.025 &  0.024 &  0.033 &  0.024 &  0.030 &  0.028 &  0.030 &  0.031 &  0.044     \\
     H38 &  18.03 &  17.68 &  17.52 &  17.27 &  17.06 &  16.99 &  16.80 &  16.71 &  16.69 &  16.61 &  16.55 &  16.46 &  16.49\\
 & 0.035 &  0.028 &  0.021 &  0.020 &  0.016 &  0.016 &  0.015 &  0.015 &  0.016 &  0.015 &  0.022 &  0.019 &  0.026     \\
     C20 &  18.44 &  18.05 &  17.87 &  17.70 &  17.50 &  17.42 &  17.24 &  17.26 &  17.13 &  17.00 &  16.90 &  16.84 &  16.81\\
 & 0.037 &  0.026 &  0.022 &  0.022 &  0.017 &  0.020 &  0.017 &  0.020 &  0.019 &  0.019 &  0.025 &  0.021 &  0.039     \\
     C38 &  18.71 &  18.35 &  18.28 &  18.05 &  17.98 &  17.88 &  17.78 &  17.70 &  17.73 &  17.73 &  17.80 &  17.80 &  17.72\\
 & 0.080 &  0.050 &  0.040 &  0.058 &  0.029 &  0.040 &  0.032 &  0.037 &  0.034 &  0.041 &  0.058 &  0.061 &  0.089     \\
     H10 &  19.27 &  18.73 &  18.54 &  18.21 &  17.96 &  17.76 &  17.53 &  17.51 &  17.31 &  17.21 &  17.19 &  17.10 &  17.03\\
 & 0.093 &  0.055 &  0.043 &  0.038 &  0.030 &  0.030 &  0.033 &  0.031 &  0.029 &  0.029 &  0.033 &  0.036 &  0.053     \\
     U137&  19.21 &  18.88 &  18.68 &  18.42 &  18.22 &  18.10 &  17.88 &  17.80 &  17.81 &  17.74 &  17.71 &  17.54 &  17.61\\
 & 0.082 &  0.058 &  0.043 &  0.043 &  0.030 &  0.030 &  0.029 &  0.031 &  0.033 &  0.036 &  0.049 &  0.042 &  0.065     \\
\hline
\end{tabular}
\end{table*}

\begin{table*}
\caption{Comparison of cluster photometry with previous measurements and the data of adopted reddening values}
\begin{tabular}{cccccccc}
\hline
\hline
Cluster & R.A. (J2000) & Decl. (J2000) & $V$ (CS1982)  & $V$ (BATC) &$B-V$ (CS1982) &$B-V$ (BATC) & $E(V-I)$ \\
(1)    & (2) & (3) & (4) & (5) & (6) & (7) & (8)\\
\hline
U49 & 01:33:45.04 & 30:47:46.7 & 16.25  & 16.21 $\pm$  0.018 & 0.68  & 0.55 $\pm$  0.032 & $0.07\pm 0.02$ \\
R12 & 01:34:07.98 & 30:38:38.0 &  ...  & 16.31 $\pm$  0.036 & 0 77 & 0.74 $\pm$  0.064 & $0.05\pm 0.03$ \\
R14 & 01:34:02.45 & 30:40:39.3 &  ...  & 16.41 $\pm$  0.035 & 0.68 & 0.86 $\pm$  0.065 & ... \\
M9 &  01:34:30.17 & 30:38:12.8 & 17.06 & 17.11 $\pm$  0.030 & 0.72 & 0.54 $\pm$  0.058 & $0.04\pm 0.03$ \\
U77 & 01:33:28.68 & 30:41:35.2 & 17.19  & 17.18 $\pm$  0.045 & 0.67  & 0.53 $\pm$  0.084 & $0.08\pm 0.03$ \\
H38 & 01:33:52.13 & 30:29:03.7 & 17.24  & 17.21 $\pm$  0.028 & 0.83  & 0.67 $\pm$  0.049 & $0.04\pm 0.01$ \\
C20 & 01:34:44.21 & 30:52:18.8 & 17.67  & 17.65 $\pm$  0.031 & 0.77  & 0.60 $\pm$  0.050 & $0.03\pm 0.03$ \\
C38 & 01:33:30.66 & 30:22:21.6 & 18.10  & 18.10 $\pm$  0.061 & 0.73  & 0.42 $\pm$  0.098 & $0.04\pm 0.02$ \\
H10 & 01:33:35.12 & 30:49:00.0 & 18.23  & 18.16 $\pm$  0.052 & 0.96  & 0.80 $\pm$  0.100 & $0.25\pm 0.03$ \\
U137 & 01:33:14.28 & 30:28:22.9 & 18.30  & 18.38 $\pm$  0.054 & 0.83  & 0.70 $\pm$  0.101 & $0.09\pm 0.03$ \\
\hline
\end{tabular}
\end{table*}

\begin{table*}
\caption{Comparison of CCD photometry (CS88) with photographic material (CS82)}
\begin{tabular}{cccc}
\hline
\hline
Cluster & $B-V$ (CS88) & $B-V$ (CS82) \\
 (1)    &     (2)      &    (3)       \\
\hline
C15 & 0.46 & 0.53 \\
C16 & 0.39 & 0.49 \\
C18 & 0.80 & 0.73 \\
C27 & 0.35 & 0.37 \\
C31 & 0.49 & 0.66 \\
U49 & 0.69 & 0.68 \\
U83 & 0.37 & 0.49 \\
U101 & 0.35 & 0.42 \\
M9  & 0.66 & 0.72 \\
\hline
\end{tabular}
\end{table*}
\begin{table*}
\caption[]{Metallicity and age distribution of ten halo globular clusters}
\begin{tabular}{cccccccc}
\hline
\hline
         &  Age~[$\log$ yr]~~~$R^2$ & Age~[$\log$ yr]~~~$R^2$ & Age~[$\log$ yr]~~~$R^2$ & Age~[$\log$ yr]~~~$R^2$ & Model of metallicity & Age~[$\log$ yr]  \\
 Cluster & $Z=0.0004$ &               $Z=0.004$ &             $Z=0.008$ &               $Z=0.02$  & adopted Z ([Fe/H])& adopted \\
\hline
      U49  & $10.30\pm 0.14$~~0.002 & $9.60\pm 0.14$~~0.001 & $9.30\pm 0.14$~~0.002 & $9.16\pm 0.14$~~0.004 & 0.004 (-0.70)& $9.60\pm 0.14$\\
      R12  & $10.30\pm 0.18$~~0.014 & $10.00\pm 0.18$~~0.001 & $9.76\pm 0.18$~~0.002 & $9.38\pm 0.18$~~0.004 & 0.004 (-0.70)& $10.00\pm 0.18$\\
      M9   & $10.30\pm 0.16$~~0.002 & $9.63\pm 0.16$~~0.001 & $9.30\pm 0.16$~~0.002 & $9.16\pm 0.16$~~0.005 & 0.004 (-0.70)& $9.63\pm 0.16$\\
      U77  & $10.15\pm 0.11$~~0.002 & $9.40\pm 0.11$~~0.001 & $9.20\pm 0.11$~~0.001 & $9.01\pm 0.11$~~0.003 & 0.008 (-0.4)& $9.20\pm 0.11$\\
      H38  & $10.30\pm 0.07$~~0.002 & $9.70\pm 0.07$~~0.001 & $9.40\pm 0.07$~~0.002 & $9.16\pm 0.07$~~0.007 & 0.004 (-0.70)& $9.70\pm 0.07$\\
      C20  & $10.30\pm 0.11$~~0.012 & $9.95\pm 0.11$~~0.003 & $9.68\pm 0.11$~~0.004 & $9.26\pm 0.11$~~0.006 & 0.004 (-0.70)& $9.95\pm 0.11$\\
      C38  & $9.28\pm 0.02$ ~~0.006 & $8.96\pm 0.02$~~0.013 & $8.81\pm 0.02$~~0.017 & $8.81\pm 0.02$~~0.020 & 0.0004 (-1.70)& $9.28\pm 0.02$\\
      H10  & $10.30\pm 0.14$~~0.009 & $9.90\pm 0.14$~~0.002 & $9.51\pm 0.14$~~0.003 & $9.26\pm 0.14$~~0.007 & 0.004 (-0.70)& $9.90\pm 0.14$\\
      U137 & $10.27\pm 0.05$~~0.003 & $9.57\pm 0.05$~~0.004 & $9.28\pm 0.05$~~0.005 & $9.16\pm 0.05$~~0.009 & 0.0004 (-1.70)& $10.27\pm 0.05$\\
\hline
\end{tabular}
\end{table*} 

{\small\hspace{-2cm}
\begin{table*}
\caption[]{Metallicity, age and reddening for R14}
\begin{tabular}{cccccc}
\hline
\hline
  Age~[$\log$ yr]$~E(B-V)$~$R^2$ & Age~~~~~$E(B-V)$~~$R^2$ & Age~~~~~$E(B-V)$~~$R^2$ & Model of metallicity & Age~[$\log$ yr]  & $E(B-V)$\\
  $Z=0.0004$ &               $Z=0.004$ &         $Z=0.02$  & adopted Z ([Fe/H])& adopted & adopted\\
\hline
 $10.30\pm 0.10$~~~0.35~~~0.004 & $9.11\pm 0.10$~~~0.38~~~0.003 & $9.23\pm 0.10$~~~0.29~~~0.005&0.004 (-0.70)
& $9.11\pm 0.10$   & 0.38 \\
\hline
\end{tabular}
\end{table*}}

\end{document}